\documentclass[11pt,twoside]{article}

\usepackage{asp2006}
\usepackage{epsf}
\usepackage{psfig}
\usepackage{lscape}

\begin{document}
\title{Hinode's SP and G-band co-alignment}   
\author{R. Centeno, B. Lites, A.G. de Wijn}   
\affil{High Altitude Observatory}    
\author{D. Elmore}
\affil{National Solar Observatory}

\begin{abstract} 
We analyze the co-alignment between Hinode's BFI-Gband images and simultaneous SP maps with the aim of characterizing the general off-sets between them and the second order non-linear effects in SP's slit scanning mechanism. We provide calibration functions and parameters to correct for the nominal pixel scales and positioning.
\end{abstract}

\section{Introduction}  

Hinode \cite{hinode, tsuneta, suematsu,  shimizu, ichimoto} hosts the first spaced-based visible Spectro-Polarimeter (SP) for high resolution observations of vector magnetic fields on the solar Photosphere. This instrument measures the full Stokes profiles of two magnetically sensitive Fe I lines on a one-dimensional slice of the solar surface at a time. The slit spans $\sim 160$\arcsec in the N-S direction, and the scanning mechanism can displace the image of the Sun at the spectrograph slit by up to $\pm 152$\arcsec, in increments of about $\sim 0.15$\arcsec.
The Focal Plane Package (FPP) of the Solar Optical Telescope (SOT)
includes broad-band and narrow-band imaging capabilities that can work simultaneously with SP. However, in order to be used simultaneously for scientific purposes, we need an accurate spatial co-alignment between both instruments.

This work aims at determining the general off-sets between SP and Gband images, as well as providing accurate values of the pixel scales.  The drift and the non-linearity of the slit scan mechanism are also characterized.

\section{Observations and method}   

Two full field of view (FOV) SP maps taken at disk center (on March 10, 2007 and October 15, 2007) with simultaneous relatively-high cadence Gband data were chosen to make a first estimate of the co-alignment between the two instruments. We used a local correlation procedure to measure, pixel by pixel, the displacement between each pair of SP and Gband images. Since the SP maps are constructed from 2047 scanning steps that take 4s each, the difference in time between the beginning (East) and the end (West) of the map is of several hours. For this reason, we cannot compare the SP map with a single Gband image, but we need to consider the closest Gband neighbor for every scanning step of SP.

Before feeding the images to the local correlation procedure we have to do some pre-processing:

\begin{itemize}
\item We first run the data through the standard reduction packages.
\item SP data consist of the full, spectrally-sampled Stokes vector. In order to compare
them to a Gband map we need to construct a continuum image from the Intensity spectral profiles.
\item Due to the time-span of a full FOV SP map and the short granulation time-scales,
in order to run a correlation with the Gband data, we need to construct a compound Gband image 
made up from slices of the Gband time series. Each slice is taken from the closest Gband neighbor
in time to the corresponding SP scanning step. 
\item We have to rescale both maps to the same pixel sizes. For this we take as a reference
the nominal scanning step, i.e. $0.1476$ \arcsec. We rescale the SP map in 
the direction along the slit to make the pixels square. Then we rescale the Gband map 
to this same pixel size and we center the images.

\end{itemize}
\subsection{Definitions}

For every position $[X_{SP}, Y_{SP}]$ in the SP image we find the one in 
the Gband image $[X_{GB},Y_{GB}]$ that yields the largest value in the
local correlation process. The displacements in the scanning direction
and along the slit ($\Delta X = X_{GB} - X_{SP}$ and $\Delta Y = {Y_{GB}-Y_{SP}}$, 
respectively) are both a function of the scanning step and the position along the slit.

\section{Analysis}
 
\begin{figure}[!h]
 \plottwo{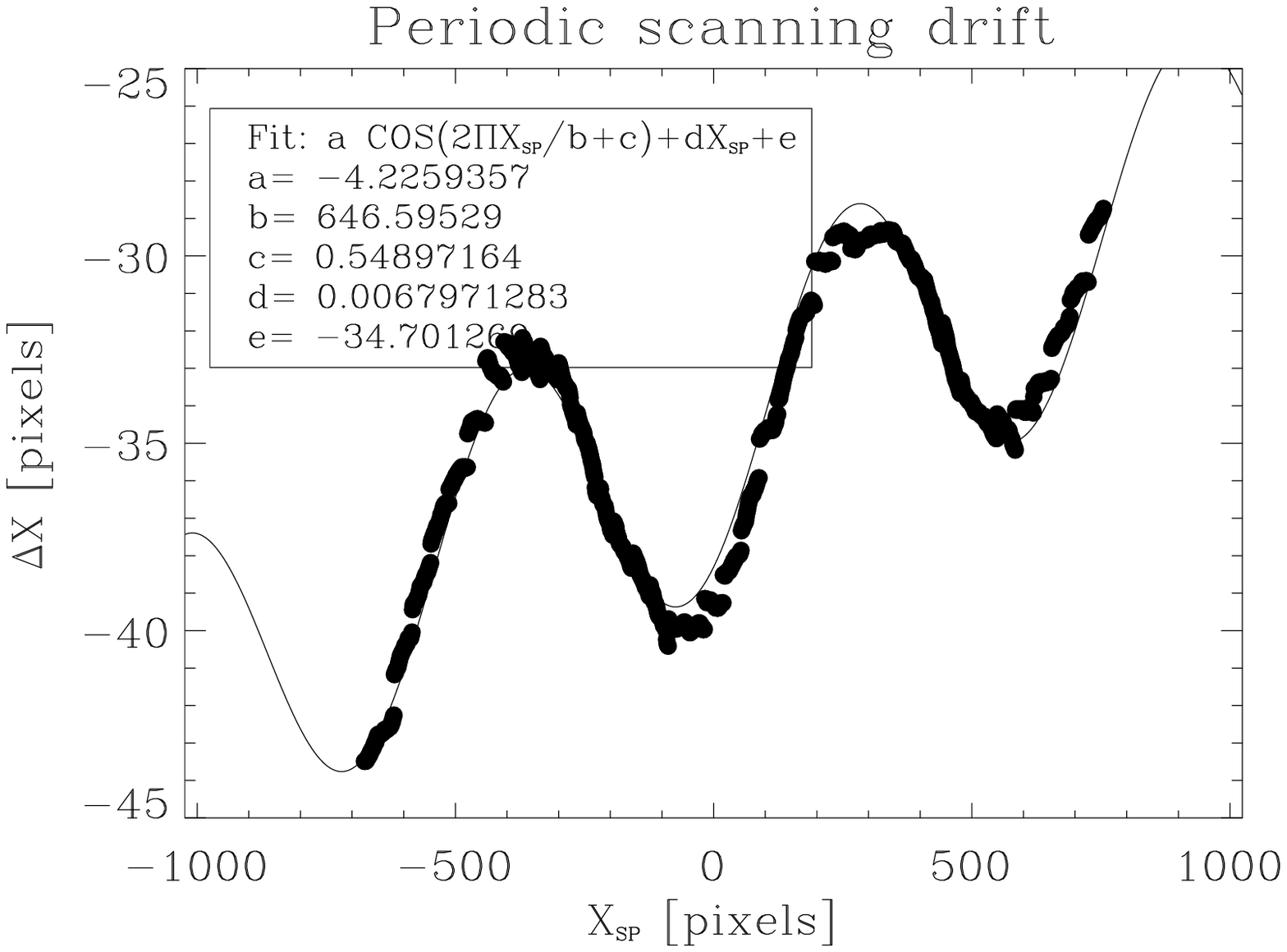}{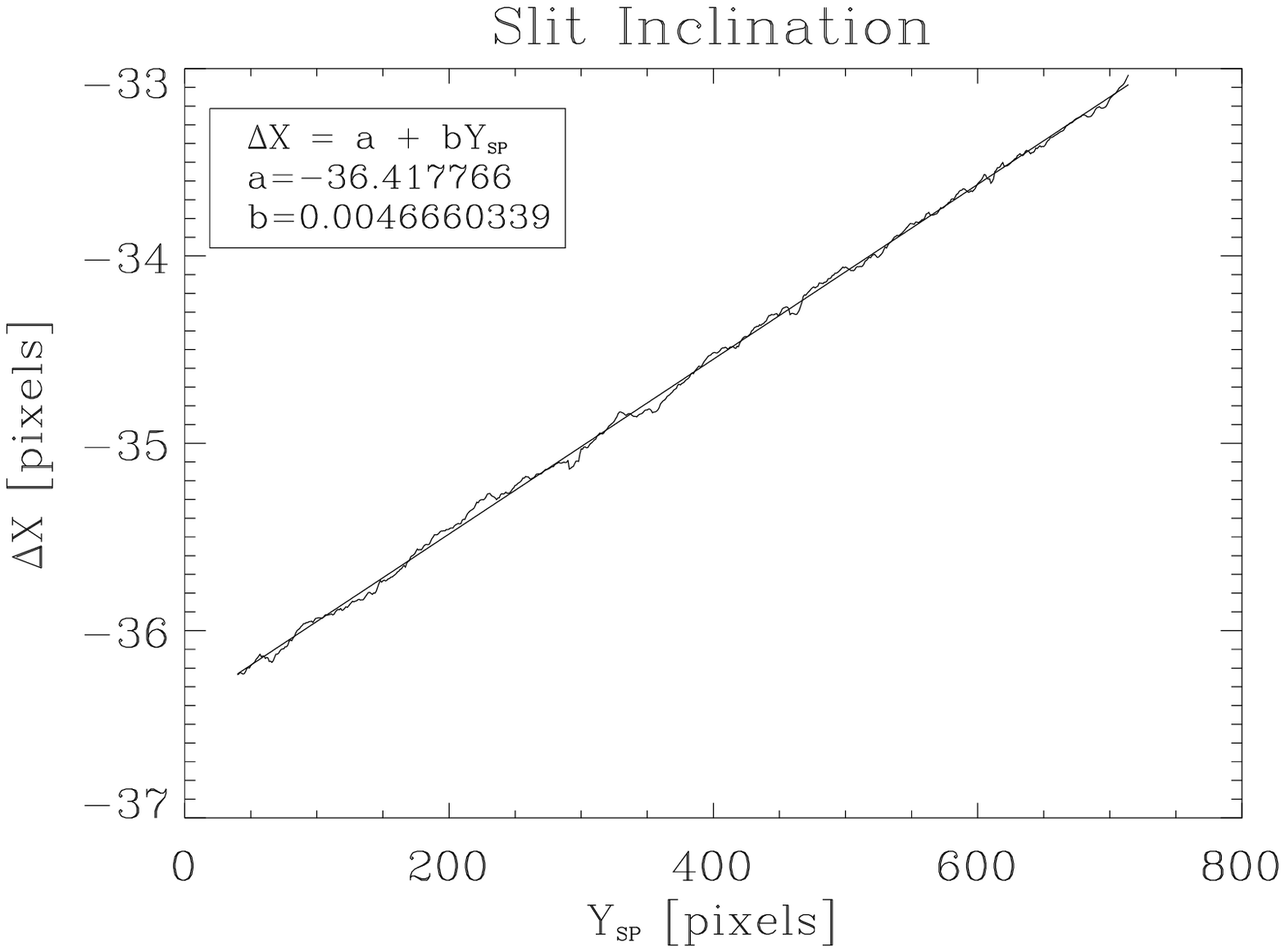}
 \plottwo{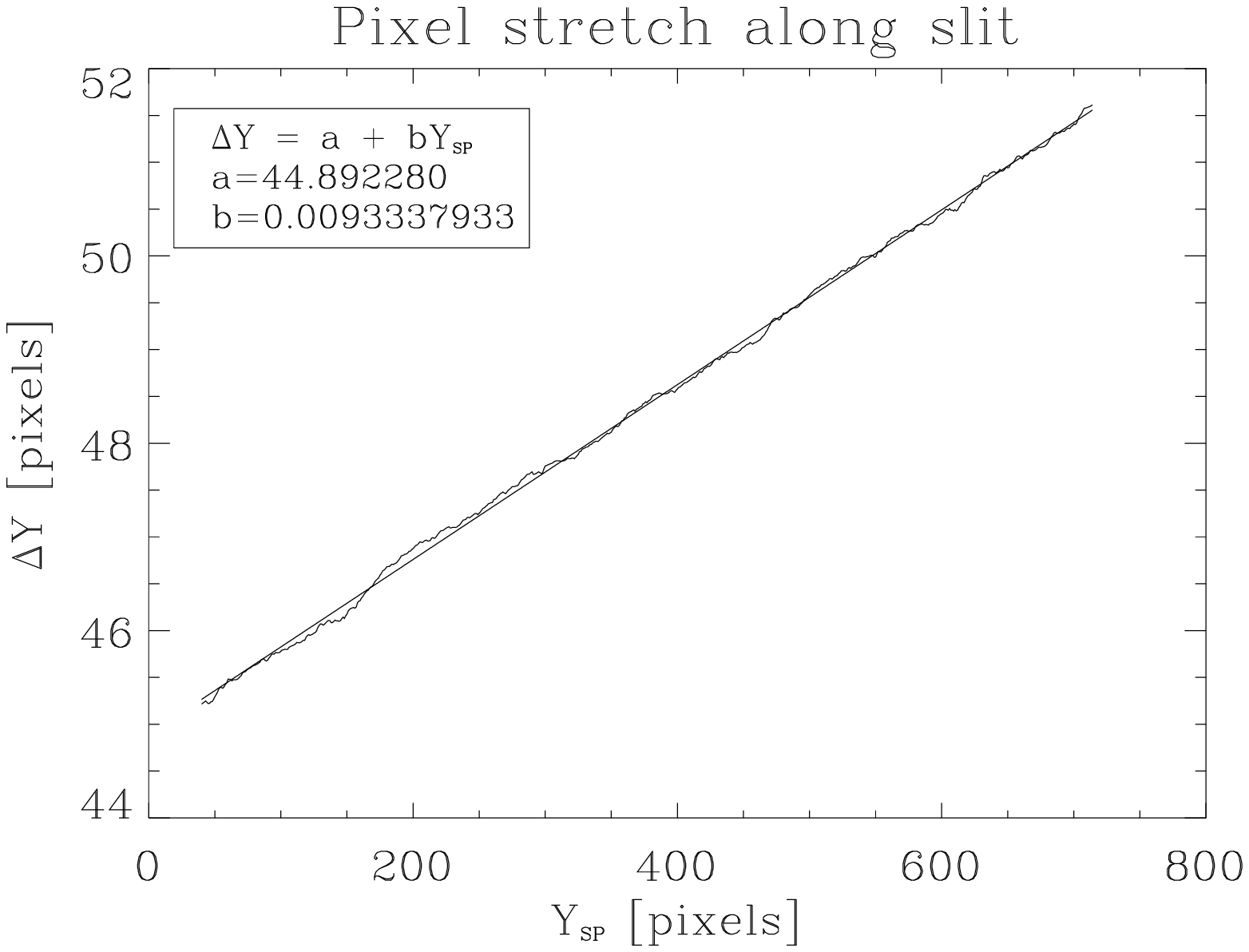}{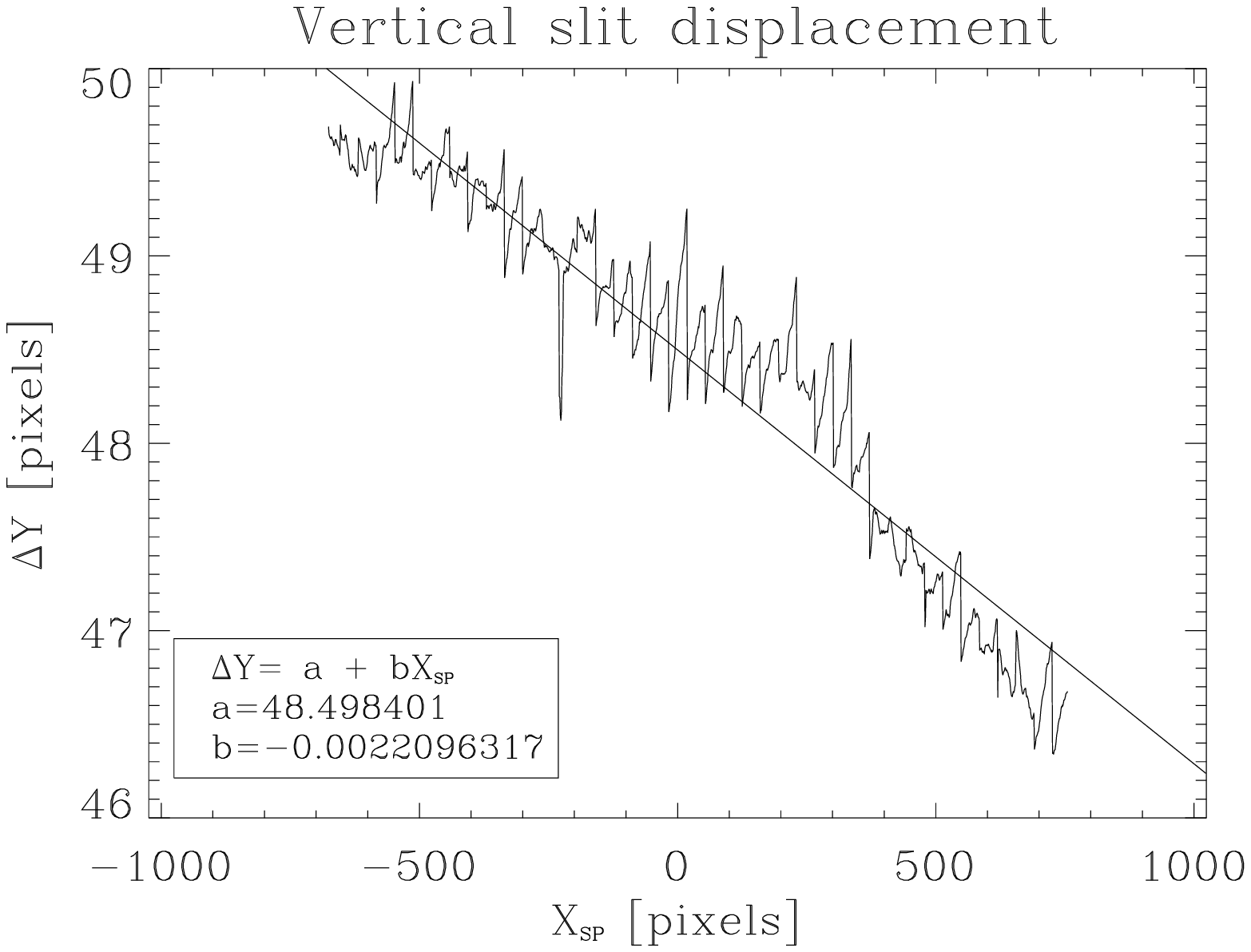}
 \caption{Average displacements, $\Delta X$ and $\Delta Y$, between the two images as a function of the position along the slit, $Y_{SP}$, and the scanning step, $X_{SP}$.\label{fig:warping}}
\end{figure}

The average displacements, $\Delta X$ and $\Delta Y$, between the two images as a function of the scanning step ($X_{SP}$) and position along 
the slit ($Y_{SP}$) are shown in Fig. \ref{fig:warping}. 

\subsection{Slit orientation}

The spectrograph slit is not oriented vertically with respect to a column
of pixels in the Gband image. The deviation is around 4 pixels along the
length of the slit. The inclination with respect to the vertical ($\alpha$) 
was obtained by fitting to a straight line ($\Delta X = a+bY_{SP}$) the average 
displacement in the scanning direction, $\Delta X$, as a function
of the position along the slit, $Y_{SP}$ (see top right panel of Fig. \ref{fig:warping}).
The inclination of the slit is given by:

\begin{equation}
\frac{\partial \Delta X}{\partial Y_{SP}} = b = tg (\alpha)
\end{equation}

\noindent where $b$ is the slope obtained from the linear fitting to the data.
SP images have to be rotated by $\alpha=0.26^{\circ}$ counter-clockwise 
with respect to the Gband images in order to correct for the inclination
of the slit.

\subsection{Vertical drift of the slit}

The bottom right panel of Fig. \ref{fig:warping} shows the averaged measured displacement of 
the slit in the vertical direction as a function of the scanning step. The linear fit of
Eq. \ref{eq:skew} yields a slope of $s = -0.0023$, indicating
that SP's slit shifts southwards with respect to the Gband image an
amount of $0.0023$ pixels for each scanning step, producing a {\em skew} effect on the image.
The off-set, $Y_{\rm{OFF}}$, gives us the bulk vertical displacement between both images.

\begin{equation}
\Delta Y = s · X_{SP} + Y_{\rm{OFF}}\label{eq:skew}
\end{equation}

\subsection{Pixel scales}

\subsubsection{Pixel scale along slit}

The average Y-displacement varies linearly as a function of the position along the slit
(see bottom left panel of the figure), implying that the nominal SP pixel size in this
direction should be corrected by a constant factor. The slope of the linear fit gives
us the correction of the pixel size (which should be increaseded by $0.94\%$ times its 
original size).

\subsubsection{Pixel scale in the scanning direction}

The average X-displacement as a function of the scanning step (top left panel of
Fig. \ref{fig:warping}) can be characterized as the sum of a cosine function and a straight 
line:

\begin{equation}          
\Delta X = A cos ( 2\pi/\Omega · X_{SP} + \delta ) + F_x · X_{SP} + X_{\rm{OFF}}\label{eq:periodic-drift}
\end{equation}

The linear trend, $F_x$, gives us information about the correction to
the size of the scanning step with respect to its nominal value, while $X_{\rm{OFF}}$
corresponds to the bulk off-set in the horizontal direction between both images. 

\noindent Due to a mechanical problem, the size of the scanning step in SP is
not a constant, but it varies periodically over a period of $\sim 650$ steps, creating
the illusion of a {\em breathing} effect on the image.
We characterized this issue using a cosine function with three free 
parameters $A$, $\Omega$ and $\delta$ (see Eq. \ref{eq:periodic-drift}), that
correspond to the amplitude of the oscillation, its period and its phase
at the origin of the reference system. 

\section{Conclusions}

We characterized the general off-sets, relative pixel scales and second order non-linear 
effects between the images produced by the SP and the Gband instruments on board the 
Hinode spacecraft.

\noindent After carrying out a preliminary calibration with the two datasets described in 
Section 2, 
we run a warping procedure over a series of 10 different datasets (with dates spanning over
more than a year), to fine-tune the calibration parameters and draw significant average values.
The final calibration parameters are compiled in the following table:

\begin{table}
\begin{tabular}{lr}
Mean scanning step ($Xscale$) :      &    $0.1486$\arcsec \\ 
Pixel scale along slit ($Yscale$) :  & $0.1599$\arcsec \\
X offset ($X_{\rm{OFF}}$) :                  & $- 4.98$\arcsec \\
Y offset ($Y_{\rm{OFF}}$) :                  & $7.26$\arcsec \\
Breathing amplitude ($A$) :          & $-0.649$\arcsec \\
Breathing period ($\Omega$) :        & $649.8$ steps \\
Breathing phase ($\delta$) :         & $0.55$ rad \\
Skew ($s$) :                         & $-0.00032$\arcsec/step \\
Slit angle :                         & $0.26 \deg$ 
\end{tabular}
\end{table}
SP's data reduction routine in the Solar Soft package (sp\_prep) uses this calibration to compute the corrected values for the FITS files header keywords XSCALE, YSCALE, XCEN and YCEN. 

\acknowledgements 

Hinode is a Japanese mission developed and launched by ISAS/JAXA, with NAOJ as domestic partner and NASA and STFC (UK) as international partners. It is operated by these agencies in co-operation with ESA and NSC (Norway). 


\end{document}